\let\ifarxiv=\iftrue     
\pdfoutput=1

\ifarxiv

\documentclass[12pt,a4paper]{article}
\usepackage[a4paper,text={450pt,650pt},centering]{geometry}

\fi

\ifarxiv\else

\documentclass[11pt,a4paper]{article}
\usepackage{mathptmx}
\usepackage[a4paper,text={130mm,198mm}]{geometry}

\fi

\usepackage{amsmath,amssymb}
\usepackage[bookmarks=true,hyperfigures=true]{hyperref}
\usepackage{graphicx}
\usepackage[nosort]{cite}
\usepackage[bulletsep]{collref}
\usepackage{multirow}
\usepackage{amsfonts}
\usepackage{amssymb}
\usepackage{amscd}

\let\oldbfseries=\bfseries
\let\oldmdseries=\mdseries
\let\oldnormalfont=\normalfont
\renewcommand{\bfseries}{\oldbfseries\boldmath}
\renewcommand{\mdseries}{\oldmdseries\unboldmath}
\renewcommand{\normalfont}{\oldnormalfont\unboldmath}

\allowdisplaybreaks[3]

\numberwithin{equation}{section}

\usepackage[font=small,labelfont=bf,width=0.85\textwidth]{caption}

\providecommand{\hypersetup}[1]{}
\providecommand{\texorpdfstring}[2]{#1}

\hypersetup{plainpages=false}
\hypersetup{pdfpagemode=UseNone}
\hypersetup{bookmarksnumbered=true}
\hypersetup{pdfstartview=FitH}
\hypersetup{colorlinks=false}
\hypersetup{citebordercolor={.5 1 .5}}
\hypersetup{urlbordercolor={.5 1 1}}
\hypersetup{linkbordercolor={1 .7 .7}}

\providecommand{\href}[2]{#2}
\providecommand{\arxivlink}[1]{\href{http://arxiv.org/abs/#1}{arxiv:#1}}

\def\IC{\mathbb{C}}
\def\IZ{\mathbb{Z}}
\def\id{{\bf 1}}

\newcommand{\Tr}{{\rm Tr}}

\def\pic #1#2{\hbox{\lower#1pt\hbox{~\mbox{\epsfxsize=20truemm \epsffile{#2}}}}}
\def\pic #1#2#3{\hbox{\lower#1pt\hbox{~\mbox{\includegraphics[scale=#3]{#2}}}}}
\def\picw #1#2#3{\hbox{\raise#1pt\hbox{~\mbox{\includegraphics[scale=#3]{#2}}}}}

\begin{document}


\thispagestyle{empty}
\phantomsection
\addcontentsline{toc}{section}{Title}

\begin{flushright}\footnotesize%
\texttt{\arxivlink{1012.4001}}\\
overview article: \texttt{\arxivlink{1012.3982}}%
\vspace{1em}%
\end{flushright}

\begingroup\parindent0pt
\begingroup\bfseries\ifarxiv\Large\else\LARGE\fi
\hypersetup{pdftitle={Review of AdS/CFT Integrability, Chapter V.1: Scattering Amplitudes -- a Brief Introduction}}%
Review of AdS/CFT Integrability, Chapter V.1:\\
Scattering Amplitudes -- a Brief Introduction
\par\endgroup
\vspace{1.5em}
\begingroup\ifarxiv\scshape\else\large\fi%
\hypersetup{pdfauthor={R. Roiban}}%
R.~Roiban
\par\endgroup
\vspace{1em}
\begingroup\itshape
Department of Physics, The Pennsylvania State University\\
University Park, PA 16802, USA
\par\endgroup
\vspace{1em}
\begingroup\ttfamily
radu@phys.psu.edu
\par\endgroup
\vspace{1.0em}
\endgroup

\begin{center}
\includegraphics[width=5cm]{TitleV1.mps}
\vspace{1.0em}
\end{center}

\paragraph{Abstract:}
We review current efficient techniques for the construction of multi-leg and multi-loop 
on-shell scattering amplitudes in supersymmetric gauge theories. Examples in the 
maximally supersymmetric Yang-Mills theory in four dimensions are included.

\ifarxiv\else
\fi
\hypersetup{pdfsubject={MSC (2010): 81T13, 81T60, 81U35}}%

\ifarxiv\else
\fi
\hypersetup{pdfkeywords={Scattering Amplitudes, Gauge Theories, Symmetry, Recursion Relations, Unitarity}}%

\newpage



\section{Introduction}

The scattering amplitudes of
on-shell excitations are perhaps the most basic quantities in any quantum field theory. 
They provide the only link between models of Nature and experimental data, being thus
an indispensable tool for testing theoretical ideas about high energy physics. They also 
contain a wealth of off-shell information, such as certain anomalous dimensions of 
composite operators, making their evaluation an important alternative approach to direct 
off-shell calculations.
 
Scattering amplitudes may exhibit larger symmetries than the 
Lagrangian\footnote{At tree-level it is possible to argue that they have the 
same symmetries as the equations of motion.}. For example, as reviewed 
in \cite{chapDual} in this volume, 
it was shown that the tree-level S-matrix 
of the ${\cal N}=4$ super-Yang-Mills theory (sYM)
is invariant \cite{Drummond:2009fd} 
under the Yangian of the four-dimensional
superconformal group, even though this is not a symmetry of the Lagrangian. 
Part of this invariance was initially observed as symmetries of
higher-loop amplitudes \cite{Drummond:2006rz}. 
Thus, in this theory, (tree-level) scattering 
amplitudes realize the symmetries responsible for the integrability of its dilatation operator
and of the worldsheet theory of its string theory dual. 
With more symmetry, one may hope that scattering amplitudes have simpler structure 
than one may naively expect.

Textbook approaches to scattering amplitude calculations make use of Feynman diagrams. 
Symmetries, however,  even those of the Lagrangian, are obscured in this approach, 
re-emerging only after all Feynman diagrams are assembled. For this reason, even at tree 
level, the evaluation of multi-leg amplitudes can become quite involved. 
Multi-loop amplitudes have similar features. Nevertheless, the fact that 
scattering amplitudes to any loop order are computable in terms of Feynman diagrams 
is an invaluable guide for identifying new techniques bypassing their 
difficulties.

Here we review the basics of modern on-shell methods for the evaluation of scattering 
amplitudes -- the (super)MHV vertex expansion, on-shell recursion relations and the
generalized unitarity-based method. Other methods and developments are briefly 
mentioned in the concluding section.

\section{Organization, presentation, relations between amplitudes}

Whether carried out in terms of Feynman diagrams or by other means, 
a good notation and a transparent organization of the calculation and
results are indispensable ingredients of an efficient calculation of scattering 
amplitudes. Color ordering separates the color flow from momentum flow and
thus separates amplitudes into smaller gauge-invariant parts -- the color-ordered 
amplitudes. Projection of these parts onto definite 
helicity configurations leads to partial amplitudes with  useful properties and
simple structure. 
An enlightening discussion of these topics may be found in \cite{Dixon:1996wi}.
Here we briefly summarize the salient points. 

\subsection{Spinor helicity and color ordering}

In a massless theory, solutions of the chiral Dirac equation 
provide an parametrization of momenta and polarization
vectors 
\cite{Berends:1981rb, DeCausmaecker:1981bg, Xu:1984qe,
Kleiss:1985yh,Gunion:1985vca,Xu:1986xb}
which allows {\it e.g.} the construction of
physical polarization vectors without fixing noncovariant gauges.
At the basis of this parametrization lies the well-known relation
\begin{eqnarray}
(k_\mu{\bar \sigma}^\mu)^{{\dot\alpha}\alpha}u_{-\alpha}(k)=0
~~;~~~
u_-(k){\bar u}_-(k)=-k_\mu {\sigma}^\mu~~,
\label{p.sigma}
\end{eqnarray}
where as usual ${\sigma}=(\id,\boldsymbol{\sigma})$ and ${\bar \sigma}
=(\id,-\boldsymbol{\sigma})$ are the
Pauli matrices .
This factorization also follows more formally from the fact that
the matrix on the right-hand-side of equation (\ref{p.sigma}) has unit rank if 
the momentum $k$ is null. It is common\footnote{In 
Minkowski signature $\lambda$ and ${\tilde \lambda}$ are complex
conjugate of each other and the factorization (\ref{p.sigma}) exhibits a 
rephasing invariance $\lambda\mapsto S\lambda,~{\tilde\lambda}\mapsto
{S^{-1}}{\tilde\lambda}$ with $S^*=S^{-1}$. 
It is useful to promote momenta to (holomorphic) complex variables and 
the Lorentz group to $SL(2,\IC)\times SL(2,\IC)$. Then, $\lambda$ and ${\tilde
\lambda}$ become independent complex variables and rephasing by $S$ becomes 
rescaling by an arbitrary complex number. Scattering amplitudes have definite 
scaling properties under this transformation.} 
to denote $u_-(k)$ and ${\bar u}_-(k)$
by $\lambda$ and ${\tilde{\lambda}}$, respectively. 
Multiplication of spinors is dictated by Lorenz invariance:
\begin{eqnarray}
\langle ij\rangle =
\epsilon^{\alpha\beta}\lambda_{i\alpha}\lambda_{j\beta} ~~~~~~~~
[ij] = -\epsilon^{{\dot \alpha}{\dot \beta}} {\tilde
\lambda}_{i{\dot \alpha}}{\tilde \lambda}_{j{\dot \beta}}~~.
\end{eqnarray}

Gauge invariance constrains the physical polarization vectors; they must 
also be transverse and take the standard form of circular polarization vectors in the 
relevant frame. They can be constructed in terms of 
$\lambda$, ${\tilde{\lambda}}$ and arbitrary fixed spinors $\xi$ and ${\tilde \xi}$:
\begin{eqnarray}
\epsilon^-_{\alpha\dot\alpha}(k, \xi)=
-\sqrt{2}\,\displaystyle{\frac{\lambda_\alpha {\tilde\xi}_{\dot\alpha}}{[\xi k]}}
~~~~~~~~
\epsilon^+_{\alpha\dot\alpha}(k, \xi)=
\hphantom{-}\sqrt{2}\,
\displaystyle{\frac{\xi_\alpha{\tilde \lambda}_{\dot\alpha}}{\langle\xi k\rangle}}~~.
\end{eqnarray}
The freedom of choosing independently reference spinors for each
of the gluons participating in the scattering process makes it easy to prove 
that tree-level gluon amplitudes with less 
than two gluons of the same helicity vanish identically. 
The first nonvanishing tree-level amplitudes have two gluons of the same helicity opposite from the other ones; they
are known as maximally helicity-violating (MHV) amplitudes. In supersymmetric theories 
this pattern holds to all orders in perturbation theory.

\

%
A clean organization of scattering amplitudes is a second useful
ingredient in the construction of scattering amplitudes at any fixed loop
order $L$.
Besides the organization following the helicity of external states,
at each loop order an organization
following the color structure is also possible and desirable, if only
because, for $n$-point amplitudes, there are at most $(n-1)!$ gauge 
invariant components. 
For an $SU(N)$ gauge theory with gauge group generators denoted by 
$T^a$, any $L$-loop amplitude may be decomposed as follows \cite{Bern:1990ux}:
\begin{eqnarray}
{\cal A}_n^{(L)}=N^L\sum_{\rho\in S_n/\IZ_n}
\Tr[T^{a_{\rho(1)}}\dots T^{a_{\rho(n)}}]A_n^{(L)}(k_{\rho(1)}\dots
k_{\rho(n)},N^{-1}) +{\rm multi{\scriptstyle -}traces}~~,
\label{color_ordering}
\end{eqnarray}
where the sum extends over all non-cyclic permutations $\rho$ of $(1\dots n)$. 
%
%
The coefficients
$A(k_{\rho(1)}\dots k_{\rho(n)},N^{-1})$ are called color-ordered
amplitudes. The $(n-1)!$ color-ordered amplitudes in (\ref{color_ordering}) are not 
independent; in \cite{DelDuca:1999rs} and \cite{BjerrumBohr:2009rd} it was shown how to express them 
in terms of $(n-2)!$ and $(n-3)!$ basic amplitudes, respectively.

In the limit of large number of colors, $N\rightarrow\infty$,
the multi-trace terms left unspecified in the equation
above drop out. The same is true for all $N$-dependent 
terms in $A_n(k_{\rho(1)}\dots k_{\rho(n)},N^{-1})$, reducing them 
to planar partial amplitudes $A_n(k_{\rho(1)}\dots k_{\rho(n)})$.
In this limit we will normalize the loop expansion parameter as
\begin{eqnarray}
a=\frac{g^2N}{8\pi^2}
\label{loop_exp_param}
\end{eqnarray}
 %
%

Color ordered scattering amplitudes have definite transformation properties under 
cyclic permutation of (subsets of) external legs. They also have definite factorization 
properties in limits in which external momenta reach certain singular configurations.
{\it E.g.} the tree-level collinear and multi-particle factorization formulae are
\begin{eqnarray}
&&\!\!\!\!\!\!\!\!\!\!\!\!\!\!\!\!\!\!\!\!
A^{(0)}_n(1 \dots (n-1)^{h_{n-1}},n^{h_n})
\stackrel{k_{n-1}|| k_n}{-\!\!\!-\!\!\!-\!\!\!\!\longrightarrow}
\sum_{h}
A^{(0)}_{n-1}(1\dots k^h){\rm Split}^{(0)}_{-h}((n-1)^{h_{n-1}},n^{h_n})~~,
\label{collinear_limit}
\\
&&\!\!\!\!\!\!\!\!\!\!\!\!\!\!\!\!\!\!\!\!
A^{(0)}_n(1,\ldots,n)\stackrel{k_{1,m}^2\rightarrow 0}{-\!\!\!-\!\!\!-\!\!\!\!
\longrightarrow}
\sum_{h= \pm} 
A^{(0)}_{m+1}(1,\ldots,m,k_{1,m}^{h}) \frac{i}{k_{1,m}^2} 
A^{(0)}_{n-m+1}(-k_{1,m}^{-h},m+1,\ldots,n)~~~~
\label{mpf}
\end{eqnarray}
where ${\rm Split}^{(0)}$ is a universal function known as the tree-level splitting amplitude.
These properties, and their higher-loop generalizations, provide stringent tests on the direct 
evaluation of higher-loop amplitudes and the validity of new methods proposed for this 
purpose.
For a thorough  discussion we refer the reader to the original 
literature \cite{Mangano:1987xk, Bern:1994zx, Kosower:1999xi}.

\subsection{Superspace and supersymmetry relations \label{susy_Ward_superspace}}

Supersymmetric field theories are more constrained than their non-supersymmetric 
counterparts. Through supersymmetric Ward identities \cite{Grisaru:1977px}, 
supersymmetry implies nontrivial relations between scattering amplitudes to all orders 
in perturbation theory.
For example, the vanishing of all gluon amplitudes with less than two gluons of 
helicity different form the rest
may be understood as a consequence of supersymmetry.
Tree-level supersymmetry relations between gluon scattering amplitudes 
hold in all theories, regardless of their amount of supersymmetry or of their field content.

Supersymmetric Ward identities imply that not all amplitudes are independent; rather, 
most of them are generated from certain "basic" amplitudes by repeated application
of supersymmetry transformations. 
{\it E.g.}, MHV amplitudes, differing by the position of the 
negative helicity gluons, are all related by supersymmetry transformations. 
The next-to-MHV amplitudes (involving three negative helicity gluons) and their 
superpartners, are generated by three independent amplitudes \cite{Broedel:2009hu, 
Elvang:2009wd}.
A general solution to the relations imposed by supersymmetry Ward identities in 
${\cal N}=4$ sYM theory and in ${\cal N}=8$ supergravity was discussed in 
\cite{Elvang:2009wd}.

Chiral superspace provides an efficient  organization of the scattering amplitudes 
of the ${\cal N}=4$ sYM theory. The physical states  are assembled into a single superfield
\begin{equation}
\Phi(x,\eta)=\frac{1}{4!}g^+_{abcd}\eta^a\eta^b\eta^c\eta^d
                    +\frac{1}{3!}f^+_{abc}\eta^a\eta^b\eta^c\
                    +\frac{1}{2!}s_{ab}\eta^a\eta^b
                    +                  f^-_a\eta^a 
                    +                  g^-
                    ~~,
\label{superfields}
\end{equation}
where $\eta$ denote the anticommuting superspace coordinates, transforming in the 
fundamental representation of the R-symmetry group $SU(4)$; $g_\pm$ and $f_\pm$ are, 
respectively, the positive and negative helicity gluons and gluinos and $s_{ab}$ are scalars. 
Component amplitudes are repackaged into superamplitudes
and can be extracted
by multiplication with a superfield containing only the desired 
component field for each external leg and integration over all 
anticommuting superspace coordinates. 

The fact that all MHV amplitudes are related by a suitable chain of supersymmetry 
transformations is reflected by the fact that all MHV amplitudes may be assembled 
into a single-term superamplitude proportional to the conservation constraint for the 
chiral supercharge 
$Q^{\alpha a}=\sum_i\lambda^\alpha_i\eta^a_i$:
\begin{equation}
{\cal A}^{(0),{\rm MHV}}_n(1, 2, \ldots, n) \equiv
\frac{i}{\prod_{j=1}^n\langle j ~(j+1)\rangle}
\, \delta^{(8)} \Bigl(\sum_{j=1}^n\lambda^j_\alpha\eta_j^a \Bigr)\, .
\label{MHVSuperAmplitude}
\end{equation}
The ${\overline {\rm MHV}}$ superamplitudes  in chiral superspace  is more 
complicated \cite{Drummond:2008bq, Bern:2009xq}:
\begin{equation}
{\cal A}^{(0),{\rm \overline{MHV}}}_n(1, 2, \ldots, n)= 
\frac{i }{\prod_{j=1}^n [ j ~(j+1)]}
\,  
\int  \prod_{a=1}^4 d^8\omega^a 
 \prod_{i=1}^n \delta^{(4)}(\eta_i^a-{\tilde\lambda}_i^{\dot\alpha}\omega_{\dot\alpha}^a)\, .
\end{equation}
Supersymmetric Ward identities imply that, to all orders in perturbation 
theory, MHV and  ${\overline {\rm MHV}}$ superamplitudes are proportional
to the corresponding tree-level superamplitude. 
The proportionality coefficient, henceforth called scalar factor and denoted by ${M}_n^{(l)}$ 
where $n$ is the number of external legs and $l$ is the loop order, is a completely symmetric 
scalar function of momentum invariants which naturally splits into parity-even and a
parity-odd components. 
The superamplitude containing the gluon amplitudes with $(k+2)$ negative 
helicity gluons (the so-called N$^k$MHV amplitudes)  contains $4(k+2)$ delta functions 
whose arguments are linear combinations of anticommuting coordinates. Examples
for $k=1$ may be found in \cite{Drummond:2008vq}.

The dual superspace, in which the superfield is related to (\ref{superfields}) by a fermionic
Fourier transform is also extensively used \cite{Drummond:2008vq,Elvang:2008na}. 
While the superamplitude is unchanged, one 
extracts component amplitudes by applying suitable fermionic differential operators.
For example, to extract a gluon amplitude one differentiates solely with respect to the $\eta$ 
parameters corresponding to the negative helicity gluons.

\subsection{Factorization of infrared divergences}

A general feature of on-shell scattering amplitudes in massless theories is the presence of 
infrared divergences.\footnote{While absent in the ${\cal N}=4$ sYM theory, in general 
massless theories ultraviolet divergences are, of course, present as well.} 
Unlike ultraviolet divergences they cannot be renormalized away; rather, they cancel
in infrared-safe quantities, such as cross sections of color-singlet states, anomalous dimesions, etc.

There are two sources of infrared divergences in a massless theory:
the small energy region of some virtual particle and the region in which 
some virtual particle is collinear with some external particle, respectively:
\begin{eqnarray}
\int \frac{d\omega}{\omega^{1+\epsilon}}\propto \frac{1}{\epsilon}
~~~~~~~~
\int \frac{d k_T}{k_T^{1+\epsilon}}\propto \frac{1}{\epsilon}~~.
\label{soft_collin}
\end{eqnarray}
Since they can occur simultaneously, the leading infrared singularity
at $L$-loops is an $1/\epsilon^{2L}$ pole in dimensional regularization.

The structure of soft and collinear singularities in a massless gauge
theory in four dimensions has been extensively studied and 
understood \cite{Akhoury:1978vq,Mueller:1979ih, Collins:1980ih, 
            Sen:1981sd,Sterman:1986aj,Botts:1989nd, Catani:1989ne, 
            Korchemsky:1988hd, Magnea:1990zb,Korchemsky:1993uz,
            Catani:1998bh, Sterman:2002qn}. 
The realization that soft and virtual collinear effects can be factorized
in a universal way, together with the fact \cite{Collins:1989gx} that 
the soft radiation can be further factorized from the (harder) collinear 
one, led to a three-factor structure for gauge theory scattering 
amplitudes \cite{Sterman:2002qn, Sen:1982bt, Aybat:2006mz}:
\begin{eqnarray}
{\cal M}_n=
    \left[\prod_{i=1}^nJ_i\left(\frac{Q}{\mu},\alpha_s(\mu),\epsilon\right)\right]
    \times S\left(k, \frac{Q}{\mu},\alpha_s(\mu),\epsilon\right)
    \times h_n\left(k, \frac{Q}{\mu},\alpha_s(\mu),\epsilon\right)~~.
\label{factorization}
\end{eqnarray}
Here the product runs over all the external lines, $Q$ is the
factorization scale, separating soft and collinear momenta, $\mu$
is the renormalization scale and $\alpha_s(\mu)=\frac{g(\mu)^2}{4\pi}$
is the running coupling at scale $\mu$. 
Both $h_n(k, {Q}/{\mu},\alpha_s(\mu),\epsilon)$ and the amplitude
${\cal M}_n$ are vectors in the space of color configurations
available for the scattering process. 
The soft function $S(k,{Q}/{\mu},\alpha_s(\mu),\epsilon)$ is a 
matrix acting on this space; it is defined up to a multiple of the 
identity matrix.  It captures the soft gluon radiation, it is responsible 
for the purely infrared poles and it can be computed in the 
eikonal approximation in which the hard partonic lines are replaced 
by Wilson lines. 
The ``jet'' functions $J_i({Q}/{\mu},\alpha_s(\mu),\epsilon)$ are color-singlets 
%
%
and contain the complete information on collinear dynamics of
virtual particles. 
Finally, $h_n(k, {Q}/{\mu},\alpha_s(\mu),\epsilon)$ contains
the effects of highly virtual fields and is finite as
$\epsilon\rightarrow 0$. The jet and soft functions can be
independently defined and evaluated in terms of specific matrix elements.

In the planar limit all except one color structure are subdominant; 
the soft function is then proportional to the identity matrix and  may 
be absorbed into the definition of the jet functions reducing equation 
(\ref{factorization}) to a two-factor expression. 
In this limit, the jet function may be given a physical interpretation
by using the factorized form of the amplitude for the decay of  a color-singlet 
state into two gluons of momenta $k_i$ and $k_{i+1}$. This is, by definition, 
the  Sudakov form factor 
${\cal M}^{[gg\rightarrow 1]}(s_{i,i+1}/\mu,\lambda(\mu),\epsilon)$. 
With this information the factorized form of a general planar amplitude
is
\begin{eqnarray}
{\cal M}_n=\left[\prod_{i=1}^n{\cal
M}^{[gg\rightarrow 1]}
\left(\frac{Q}{\mu},\lambda(\mu),\epsilon\right)\right]^{1/2}
    \times 
    h_n\left(k, \frac{Q}{\mu},\lambda(\mu),\epsilon\right)~~,
\label{factorization_sudakov}
\end{eqnarray}
where $\lambda(\mu)=g(\mu)^2N$ is the 't~Hooft coupling. 
Here ${\cal M}_n$ denotes the unique single-trace structure relevant in the 
planar limit.

Independence on the factorization scale $Q$ implies that the Sudakov form factor
obeys certain renormalization group type equations which relate it to the cusp 
anomalous dimension as well as to another function | the "collinear anomalous 
dimension" | whose physical  interpretation is less transparent (see 
however \cite{Dixon:2008gr}).
For their derivation and analysis we shall refer the reader to the original 
literature \cite{Mueller:1979ih,Collins:1980ih,Sen:1981sd,Ivanov:1985np,
Korchemsky:1985xj}. Their solution for ${\cal N}=4$ sYM is \cite{Bern:2005iz}:
\begin{eqnarray}
{\cal M}_n&=& \exp\left[-\frac{1}{8}\sum_{l=1}^\infty
a^l \left(\frac{\gamma_K^{(l)}}{(l\epsilon)^2}
     +\frac{2{\cal G}_0^{(l)}}{l\epsilon}\right)\sum_{i=1}^n
\left(\frac{\mu^2}{-s_{i,i+1}}\right)^{l\epsilon}\right]\;\times\;h_n~~,
\label{soft_collinear}
\end{eqnarray}
where the cusp anomaly (universal scaling function) and the collinear 
anomalous dimension are constructed from the coefficients $\gamma_K^{(l)}$ \
and ${\cal G}_0^{(l)}$  as:
\begin{eqnarray}
f(\lambda)\equiv \gamma_K(\lambda)=\sum_{l} a^l \gamma_K^{(l)}
~~~~~~~~
G_0=\sum_l {\cal G}_0^{(l)} a^l~~.
\label{coefs}
\end{eqnarray}
In writing (\ref{soft_collinear}) it was assumed that the factorization scale of IR divergences 
associated to the external legs carrying momenta $k_i$ and $k_{i+1}$ is $Q=s_{i.,i+1}$.

The detailed structure of IR divergences of scattering amplitudes described above used 
to great effect \cite{Bern:2006ew} for the evaluation of the 4-loop cusp anomaly which 
tests the detailed structure of the BES equation \cite{Beisert:2006ez} and thus of 
integrability for ${\cal N}=4$ sYM theory. The BES equation provides all-order
results for $\gamma_K$; no such all-order determination of 
the collinear anomalous dimension is available, though its relation to other anomalous 
dimensions \cite{Dixon:2008gr} may remedy this situation.

\section{Tree level amplitudes}

All symmetries of the Lagrangian of a quantum field theory are visible in its on-shell 
scattering amplitudes. Scattering amplitudes may however have more symmetries than 
the Lagrangian. New presentations of scattering amplitudes may thus expose hitherto 
unsuspected hidden properties of the theory.

An enigmatic presentation of tree-level scattering superamplitudes of ${\cal N}=4$ sYM 
followed \cite{Roiban:2004yf} from Witten's interpretation 
of the theory as a topological string theory in the super-twistor
space of super-Minkowski  space. The
generating function of tree-level amplitudes with $n$ external legs is
\begin{eqnarray}
A_n=\sum_{d=2}^{n-3}\int d{\cal M}_{1,d} \, \langle J_1\dots J_n\rangle
\end{eqnarray}
where $d{\cal M}_{1,d}$ is the integration measure over the moduli
space of maps of degree $d$ from $S^2$ to $CP^{3|4}$ and $J_i$ are 
certain free fermion currents.
Recently, the properties of this presentation of amplitudes started being 
understood
\cite{ArkaniHamed:2009dn,ArkaniHamed:2009vw,Mason:2009qx,
Spradlin:2009qr,Nandan:2009cc}
through the Grassmannian interpretation of the tree-level amplitudes.

Witten's interpretation of ${\cal N}=4$ sYM  theory as a topological string theory also 
led to the MHV vertex rules \cite{Cachazo:2004kj} subsequently
generalized to the super-MHV vertex rules.\footnote{The MHV (super)vertex rules 
were proven from a Lagrangian standpoint in 
\cite{Gorsky:2005sf, Mansfield:2005yd, Ettle:2008ey,
Elvang:2008vz, Boels:2006ir, Feng:2006yy}.} 
They are effective rules expressing general amplitudes as sums of products of MHV 
superamplitudes. 
The following (super)steps generate the $n$-point N$^k$MHV gauge
theory superamplitude:

\begin{itemize}

\item draw all
tree graphs with $(k+1)$ vertices, on which the $n$ external legs are
distributed in all possible inequivalent ways while maintaining the
color order.  

\item
 to each vertex associate an MHV superamplitude
(\ref{MHVSuperAmplitude}).  The
holomorphic spinor $\lambda_P$ 
of an internal line is
constructed from the off-shell momentum $P$ of that 
line using a fixed arbitrary 
reference anti-holomorphic spinor $\zeta^{\dot\alpha}$:
\begin{equation} 
\lambda_{P\alpha} \equiv
P_{\alpha{\dot\alpha}}\zeta ^{\dot\alpha} \,.
\label{CSW_spinor}
\end{equation} 
Alternatively, the holomorphic spinor $\lambda_P=|P^\flat\rangle$
is constructed from the null projection of the off-shell momentum $P$ 
along a reference null vector $\zeta^\mu$ common for all 
legs \cite{Kosower:2004yz, Bena:2004ry}:
\begin{equation}
P^\flat = P -\frac{P^2}{2\zeta\cdot P }\zeta\,.
\label{CSWMomShift}
\end{equation}
%

\item
to each internal line 
associate a super-propagator, {\it i.e.} 
a standard 
scalar Feynman propagator $i/P^2$ and a factor which equates the 
fermionic coordinates $\eta$ of the internal line in the two vertices
connected by it.\footnote{For the superfields (\ref{superfields}) this factor 
is just  $\int d^4\eta' \delta^{(4)}(\eta-\eta')$.} 

\item 
integrate over all the anticommuting coordinates associated to internal lines.

\end{itemize}

\noindent
Upon application of these rules, the N$^k$MHV  superamplitude is given by
\begin{equation}
{\cal A}^{{\rm N}^k{\rm MHV}}_n = 
i^m \sum_{\rm all~graphs}\int \Bigl[ \prod_{j=1}^{k} {d^4\eta_j\over P^2_j}\Bigr]
 {\cal A}^{\rm MHV}_{(1)} {\cal A}^{\rm MHV}_{(2)}\cdots 
  {\cal A}^{\rm MHV}_{(k)}{\cal A}^{\rm MHV}_{(k+1)}\,,
\label{CSWequation}
\end{equation}
where the integral is over the $4k$  internal Grassmann parameters 
($d^4\eta_j \equiv \prod_{a=1}^4d\eta_j^a$) 
and each $P_j$ is the off-shell momentum of the $j$'th internal leg of the 
graph. 

Each integration $\int d^4\eta_i$ in (\ref{CSWequation}) selects the
configurations with exactly four distinct $\eta$-variables
$\eta^1_i\eta^2_i\eta^3_i\eta^4_i$ on each of the internal lines.
Since a particular $\eta^a_i$ can originate from either of two MHV
amplitudes connected by the internal line $i$, 
there are $2^4$ possibilities that may give non-vanishing contributions. 
However, for a given choice of external states, each term
corresponding to a distinct graph in (\ref{CSWequation}) receives
nonzero contributions from exactly one state for each internal leg.

The observation that integrating over the common
$\eta$ variables  yields a sum over the 16 states in the $N=4$ multiplet 
will be important also in the following 
sections in evaluating similar sums (called "supersums") appearing in generalized 
unitarity cuts.  

The simplest example illustrating the MHV (super)vertex rules is the 
construction of  the ${\overline{\rm MHV}}$ gluon amplitude; its split
helicity configuration is simply:
\begin{eqnarray}
&&\!\!\!\!\!\!\!\!\!\!\!\!
A_5^{(0)}{\scriptstyle{(1^-,2^-,3^-,4^+,5^+)}}=
\\
&&\!\!\!\!\!\!\!\!
\frac{\langle 2 3\rangle^4 }{\langle 2 3 \rangle  \langle 3 4 \rangle 
\langle 4 P_1\rangle\langle P_1 2\rangle}
\frac{1}{P_1^2}
\frac{\langle 1 P_1\rangle ^4 }{\langle 5 1\rangle\langle 1 P_1\rangle\langle P_1 5\rangle}
+
\frac{\langle 3 P_2\rangle^4}{\langle P_2 3 \rangle  \langle 3 4 \rangle 
\langle 4 P_2\rangle}
\frac{1}{P_2^2}
\frac{\langle 1 2\rangle^4}{\langle 2 P_2\rangle \langle P_2 5 \rangle
              \langle 5 1 \rangle\langle 1 2\rangle}\cr
&&\!\!\!\!\!\!\!\!\!\!\!\!
+
\frac{\langle 3 P_3\rangle^4}{\langle 3 4\rangle  \langle  45 \rangle 
\langle 5 P_3\rangle\langle P_3 3 \rangle}
\frac{1}{P_3^2}
\frac{\langle 1 2\rangle^4}{\langle  P_3 1\rangle \langle 1 2\rangle 
               \langle 2 P_3\rangle}
+
\frac{\langle 2 3\rangle ^4}{\langle 2 3\rangle  \langle  3 P_4 \rangle 
\langle P_4 2\rangle}
\frac{1}{P_4^2}
\frac{\langle 1 P_4\rangle }{\langle  1P_4\rangle \langle P_4 4\rangle 
               \langle 45\rangle\langle  5 1\rangle}
\nonumber
\end{eqnarray}
The momenta $P_i$ follow from momentum conservation at each MHV vertex;
their null components assumed above are obtained as in (\ref{CSWMomShift}).

While much more efficient than Feynman diagrams, the MHV supervertex expansion
is not recursive and the number of contributing graphs grows quite fast with the number 
of external legs; it also exhibits an artificial lack of covariance at intermediate stages due to
the presence of the fixed spinors $\zeta$.  
%
The BCFW recursion relation \cite{Britto:2005fq}
reconstruct covariantly tree-level amplitudes from this pole structure
and their multi-particle factorization properties. 

Their direct derivation \cite{Britto:2005fq} uses only complex analysis.
One singles out two momenta $p_i$ and $p_j$ (the choice of momenta 
is, to a large extent, arbitrary; we will discuss shortly the origin
of constraints on the choice of $i$ and $j$) and shifts them as
\begin{eqnarray}
p_i\rightarrow {\hat p}_i=p_i+z{\zeta}_{ij}
~~~~~~~~
p_j\rightarrow {\hat p}_j=p_j-z{\zeta}_{ij}
~~~~~~~~
(\zeta{}_{ij}){}_{\alpha{\dot\alpha}}=
\lambda_{i\alpha}{\tilde \lambda}_{j{\dot\alpha}}
\label{BCFWshift}
\end{eqnarray}
where the vector $(\zeta{}_{ij})$  is chosen such that 
the shifted momenta are still null. More elaborate shifts have also been discussed.
By tuning the parameter $z$ it is possible to expose one by one all 
poles of the amplitude. 
As the relevant values of $z$ are complex, equation~(\ref{BCFWshift}) is 
interpreted as an analytic continuation to complex momenta.

The fact that the only poles of the shifted amplitude arise from the $z$ 
dependence of propagators implies that none of them is at $z=0$.
\footnote{Poles on the $z$-plane may drift close to the origin only in
multi-particle factorization limits of the {\it unshifted} amplitude.}
The original (unshifted) amplitude may them be recovered by integrating
the shifted amplitude on a small contour $C_0$ around $z=0$.
Reinterpreting it as a contour around $z=\infty$ implies that the
amplitude may be rewritten in terms of the residues of the shifted 
amplitude. Since the corresponding poles are in one to one
correspondence with multi-particle factorization limits of the shifted
amplitude, it follows from eq.~(\ref{mpf})  that their residues are
themselves products of amplitudes. We are finally led to
\cite{Britto:2005fq} 
\begin{eqnarray}
A{\scriptstyle(1\dots n)}&=&
\frac{1}{2\pi i}
\oint_{C_0} \frac{dz}{z}A{\scriptstyle({\hat 1},2\dots {\hat n};z)}
=
-\frac{1}{2\pi i}
\oint_{C_\infty} \frac{dz}{z}A{\scriptstyle({\hat 1},2\dots {\hat n};z)}\cr
&=&
\sum_{l,h}
      A_L{\scriptstyle{({\hat 1},2\dots l, {\hat q}{}^h ;z_{0l})}}
	\frac{1}{P_{1,\dots ,l}^2}
      A_R{\scriptstyle{(-{\hat q}{}^{-h},(l+1),\dots {\hat n};z_{0l})}}
    +{\cal C}_{\infty}~~,
\label{RR_gen}
\end{eqnarray}
where $h$ denotes the helicity of the intermediate leg.
For definiteness and ease of notation we chose to shift the external
momenta $p_1$ and $p_n$; the momentum ${\hat q}$ of the internal line 
is determined by momentum conservation and depends on ${z}$.
The value of $z_{0l}$ is determined from the on-shell condition for
the intermediate line:
\begin{eqnarray}
z_{0l}=\frac{P_{1,\dots ,l}^2}{2\, {\zeta}_{1n}\cdot P_{1,\dots ,l}}~~.
\end{eqnarray}
The term denoted by ${\cal C}_{\infty}$ represents the contribution of
the pole at $z=\infty$. It is possible to argue \cite{Britto:2005fq} using
either Feynman diagrammatics or the MHV vertex rules that this
contribution is absent for the shift (\ref{BCFWshift}) for all choices of
helicity for the legs $(i,j)$ except $(h_i,h_j)=(+,-)$. 
%

Some of the terms in the sum in equation (\ref{RR_gen}) contain
three-particle amplitudes. The analytic continuation to complex momenta 
({\it i.e.} $\lambda\ne (\tilde\lambda){}^*$)
makes these terms nonvanishing.\footnote{
Either the MHV or the ${\overline{\rm MHV}}$ three-particle amplitude may be chosen 
nonvanishing, but not both.} 

\begin{figure}[t]
\centerline{\includegraphics[width=4.6in]{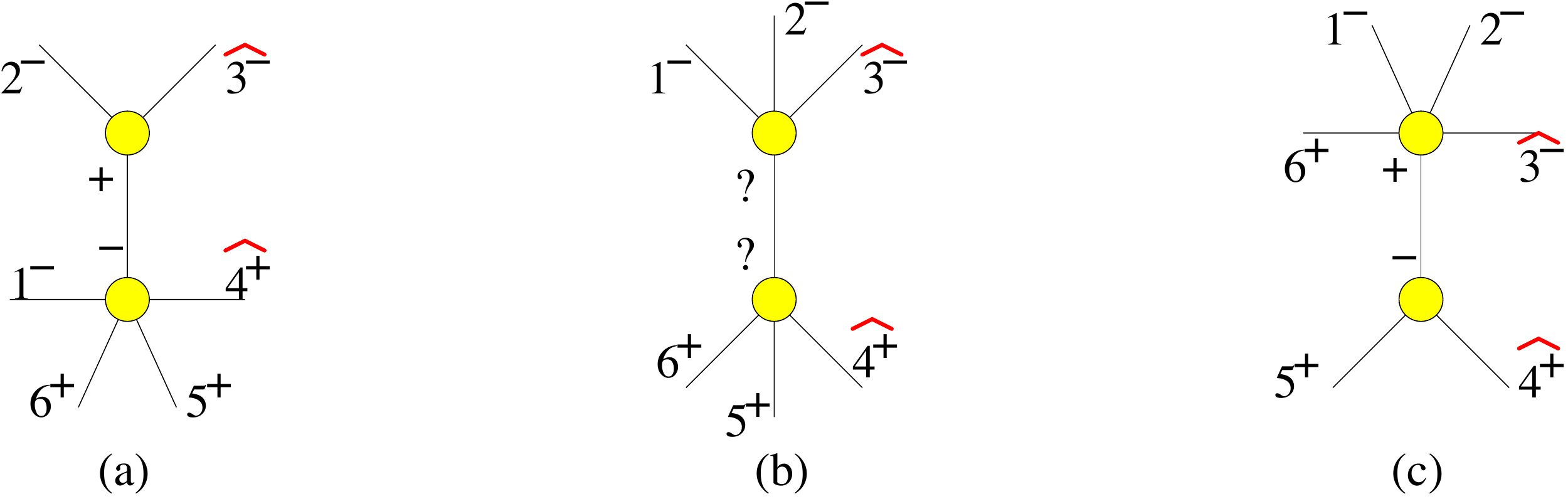}}
\caption[a]{\small The diagrammatic presentation of the terms in
equation (\ref{RR_gen}) for $A{\scriptstyle(1^-2^-3^-4^+5^+6^+)}$.}
\label{fig:RecRel}
\end{figure}

The six-point amplitude in split helicity configuration 
$A{\scriptstyle(1^-2^-3^-4^+5^+6^+)}$ provides a simple illustration 
of the BCFW recursion relations. Choosing to shift 
the momenta $p_3$ and $p_4$, the diagrams representing the terms
in equation (\ref{RR_gen}) are shown in figure~\ref{fig:RecRel}.
Diagram $(b)$ vanishes identically; the other two contribute as follows:
\begin{eqnarray}
T_a&=&\frac{\langle 2\widehat{3}\rangle^3}{\langle\widehat{3}\widehat{p_{23}}\rangle
\langle \widehat{p_{23}} 2\rangle}\,\frac{1}{p_{23}^2}\,
\frac{\langle1\widehat{p_{23}}\rangle^3}
{\langle\widehat{p_{23}}\widehat{4}\rangle\langle\widehat{4}5\rangle\langle56\rangle
\langle61\rangle}~~~~~~~~~~~~~~z_{01}=\frac{p_{23}^2}{\langle
4|P_{23}|3]}\nonumber\\[5pt]
T_c&=&\frac{[\widehat{p_{45}}6]^3}
{[\widehat{p_{23}}6][61][12][2\widehat{3}][\hat{3}
\widehat{p_{23}}]}\,\frac{1}{p_{45}^2}\,
\frac{[\widehat{4}5]^3}{[5\widehat{p_{45}}][\widehat{p_{45}}
\widehat{4}]}
~~~~~~~~~~z_{03}=\frac{p_{45}^2}{\langle4|p_{45}|3]}
\end{eqnarray}
Combining them and making use of the corresponding shifts leads to
\begin{eqnarray}
A{\scriptstyle(1^-2^-3^-4^+5^+6^+)}=\frac{1}{\langle5|p_{34}|2]\vphantom{|^{|}}}
\left[
\frac{\langle1|p_{23}|4]^3}{[23][34]\langle56\rangle
\langle61\rangle p_{234}^2\vphantom{|^{|{}^a}}}+
\frac{\langle3|p_{45}|6]^3}{[61][12]\langle34\rangle
\langle45\rangle p_{345}^2\vphantom{|^{|{}^a}}}
\right]~~.
\end{eqnarray}
This is indeed the correct answer for the six-point split-helicity
tree-level gluon amplitude, as may be verified by direct comparison
with the classic results of \cite{Mangano:1987xk}.

BCFW recursion relations have been generalized
\cite{Brandhuber:2008pf, ArkaniHamed:2008gz, Elvang:2008na} to chiral
on-shell superspace. By solving them explicit expressions for all
tree-level amplitudes of ${\cal N}=4$ sYM have been obtained in \cite{Drummond:2008cr}.


\section{Generalized unitarity and loop amplitudes} 

As explained in the previous section, the MHV vertex rules and 
the on-shell recursion relations may be understood as procedures for 
reconstructing a function of many variables from its singularities
and behavior at infinity.

Historically, through the optical theorem,
such a strategy was first used to construct loop amplitudes. 
Unitarity of the scattering matrix implies that its interaction
part $S=\id+iT$ obeys the equation:
\begin{eqnarray}
i(T^\dagger-T)=T^\dagger T~~.
\label{unitarity_V0}
\end{eqnarray}
Expanding both sides in the coupling constant implies that, at
loop order $L$, the discontinuity\footnote{This interpretation 
is a consequence of the $i\epsilon$ prescription: 
$\frac{1}{l^2+i\epsilon}-\frac{1}{l^2-i\epsilon}=  -2\pi i \theta(l^0)\delta(l^2)$.}  
-- or cut -- of $T$ in some multi-particle invariant is given by the
product of lower order terms in the perturbative expansion of 
the $T$ matrix, i.e. lower order on-shell amplitudes.

For bookkeeping purposes it is useful to separate cuts in two classes:
singlet and non-singlet. In the former only one type of field
crosses the cut. In the latter several types of particles --
complete multiplets in a supersymmetric theory -- cross the cut.
The summation over all such states can be tedious; at low orders
it may be explicitly carried out using the component version of the
supersymmetric Ward identities. General procedures, based on chiral superspace,
for effortlessly carrying out such sums -- called supersums -- have been described in 
detail in \cite{Elvang:2008na, Bern:2009xq}.

Reconstructing an amplitude from its unitarity cuts is not completely 
straightforward. One of the main difficulties is that the emerging 
integrals -- dispersion integrals -- are {\it not} of  the type usually found in 
Feynman diagram calculations.
A reinterpretation of the equation (\ref{unitarity_V0}) bypasses this
issue, expresses the result in terms of Feynman integrals  and allows 
use of the recent sophisticated techniques for their evaluation:
integral identities, modern reduction techniques, differential
equations, reduction to master integrals, etc.

To reinterpret the $L$-loop component of eq.~(\ref{unitarity_V0}) 
we notice that, due to the Feynman diagrammatics  underlying the 
amplitude calculation, it is possible to identify on the left-hand side 
of this equation all the terms with a prescribed set of cut propagators.
Equation~(\ref{unitarity_V0}) expresses the sum of these terms as a 
product of lower-loop amplitudes. Thus,  at the level of the amplitudes' 
integrand, a unitarity cut may be interpreted as isolating the terms containing 
a prescribed set of (cut) propagators.

These observations, originally due to Bern, Dixon, Dunbar and Kosower
\cite{Bern:1994zx} and improved at one-loop level by
Britto, Cachazo and Feng \cite{Britto:2004nc},
allow ``cutting'' more than $(L+1)$ propagators for an $L$-loop
amplitude, generalizing the unitarity relation (\ref{unitarity_V0}). 
These generalized cuts\footnote{Similarly to regular cuts, generalized cuts can
be either of singlet and non-singlet types.} 
do not have the interpretation of the imaginary part 
of some higher-loop amplitude. Rather, they should be interpreted as 
isolating the terms that contain a prescribed set of propagators.
The Feynman rules underlying the calculation guarantee that the 
totality of generalized cuts contains the complete information necessary to 
reconstruct the amplitude to any order in perturbation theory.
Indeed, each term in the integrand of the amplitude contains (perhaps 
after integral reduction) some subset of 
the propagators required by Feynman rules and each such term is captured by 
at least one generalized cut.

These arguments assume that the generalized cuts are constructed
in the regularized theory. In the following dimensional regularization with 
$d=4-2\epsilon$ is assumed. \footnote{In planar 
${\cal N}=4$ sYM specific patterns of breaking of gauge symmetry also
provide successful IR regularization \cite{Alday:2009zm}. We will comment 
on their features in the concluding section.}
 In practice it is convenient to start by analyzing four-dimensional
 cuts, as one can saturate them with four-dimensional helicity states
 and also make use 
 of the supersymmetric Ward identities.
The terms  arising from the $(-2\epsilon)$-dimensional components 
of the momenta in momentum-dependent vertices that are potentially
missed by four-dimensional cuts are separately found either
by a comparison with $d$-dimensional cuts or by other means.
%
%
In supersymmetric theories it can be argued \cite{Bern:1994cg}
based on the improved power-counting of the theory that, through 
${\cal O}(\epsilon^0)$, one-loop amplitudes follow from
four-dimensional cuts.

In \cite{Bern:2010qa} a generalized unitarity approach was proposed for 
theories that may be continued to six dimensions. This construction, which 
is based on a six-dimensional version of spinor helicity 
\cite{Cheung:2009dc,Dennen:2009vk}, 
provides 
a natural context form the ${\cal O}(\epsilon)$ components of momenta
and allows a Coulomb-branch regularization of IR divergences.

An $L$-loop $n$-point amplitude has (very) many generalized cuts; it is 
important to evaluate them such that the maximum number of terms is 
determined with the least amount of effort. A strategy initially advocated 
in \cite{Bern:2007ct} and extensively used in \cite{Bern:2008pv,Bern:2010tq}
is to begin with the generalized cuts imposing $4L$ cut conditions
(maximal cuts) and then proceed by releasing the on-shell condition
for one propagator at a time (near-maximal cuts). 
This is known as "the method of maximal cuts".


\section{One loop amplitudes}

\begin{figure}[t]
\centerline{
\includegraphics[width=3.5in]{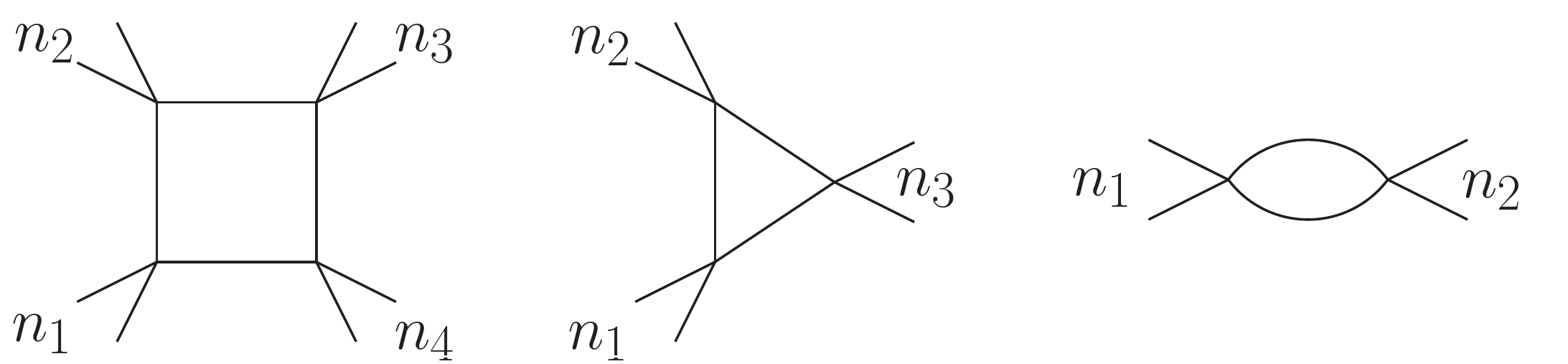}
}
\caption[a]{\small Box, triangle and bubble integrals with arbitary
numbers of external legs $n_{1,2,3,4}$ at each vertex.}
\label{fig:btb}
\end{figure}

%
Quite generally in four dimensions, such one-loop scattering amplitudes in a massless 
supersymmetric theory may be shown to be a linear combination of 
scalar box, triangle and bubble integrals (see Figure \ref{fig:btb})
with coefficients depending on the external
momenta. \footnote{Non-supersymmetric theories contain additional
rational terms. Their determination is beyond the scope of this
review. See however \cite{Berger:2009zb} and references therein.}
In ${\cal N}=4$ sYM it is possible to argue \cite{Bern:1994zx}
that amplitudes with external states belong to the same
${\cal N}=1$ vector multiplet may be written as a sum of box
integrals:\footnote{ The box integrals, represented graphically 
in Figure~\ref{fig:btb}(a), are defined and given in reference
\cite{Bern:1992em,Bern:1993kr} (with the four-mass boxes from
ref. \cite{Denner:1991qq,Usyukina:1992jd,Usyukina:1993ch}).}
\begin{eqnarray}
A_{n}^{(1)}=\sum_{ijk} c_{ijkl}I_{ijkl}~~.
\label{sum_box}
\end{eqnarray}
Experience shows that the same holds for other external states as
well. In eq.~(\ref{sum_box}) $(i,j,k,l)$ are cyclic labels of the
first external leg  at each corner of the box (counting clockwise), 
$I_{ijkl}$ is the corresponding integral 
and the sum runs over all ways of choosing the labels $(i,j,k,l)$. These integrals 
are linearly independent (over rational, momentum dependent coefficients) so this 
decomposition is unique.

Since each box integral has an unique set of four propagators, 
a quadruple cut 
(i.e. the result of eliminating four propagators and using the on-shell 
condition for their momenta)
isolates an unique box integral and its coefficient \cite{Britto:2004nc}.
Following the previous discussion, the quadruple cut of the amplitude
is simply given by the product of four tree amplitudes evaluated on
the solution of the on-shell conditions for the four propagators:
%
\begin{eqnarray}
c_{ijkl}=\frac{1}{2}\sum_{h_{q_i}}
A{\scriptstyle (q_1,i\dots j-1,-q_2)}A{\scriptstyle (q_2,j\dots k-1,-q_3)}
A{\scriptstyle (q_3,k\dots l-1,-q_4)}A{\scriptstyle (q_4,l\dots
i-1,-q_1)}
\Big|_{q_i^2=0}
\label{qc_coefs}
\end{eqnarray}
The sum runs over all possible helicity assignments on the internal lines.
The factor of $1/2$ above is due to the four on-shell conditions
having two solutions with equal values of the quadruple-cut box
integrals. The sum over these solutions is implicit in the
sum in equation (\ref{qc_coefs}).
It is important to realize that any amplitude contains at least one
box integral with one three-point corner. To construct its coefficient
through this method it is necessary to analytically 
continue momenta to complex values.

\begin{figure}[t]
\centerline{
\includegraphics[width=5.0in]{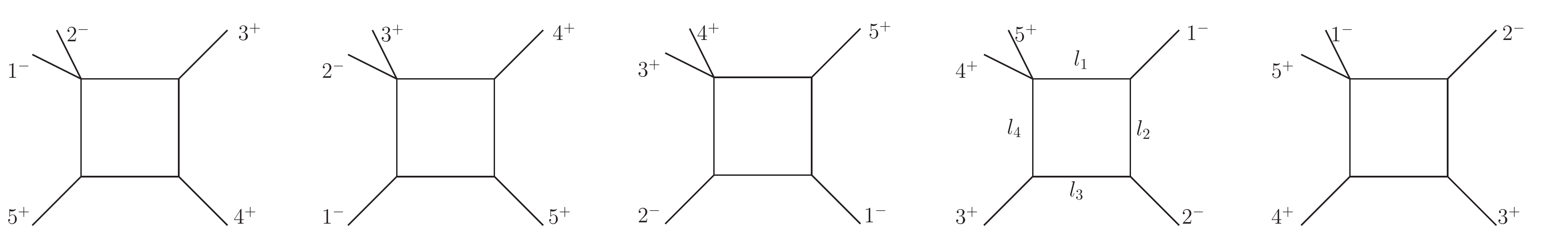}
}
\caption[a]{\small Contributions to the one-loop five-point MHV amplitude.}
\label{fig:5pt_1loop}
\end{figure}

The calculation of the five-point amplitude, initially computed
by other means \cite{Bern:1993mq} in both in ${\cal N}=4$ sYM and QCD,
is a simple illustration of the quadruple cut approach.
The five possible integral contributions are shown in Figure~\ref{fig:5pt_1loop}. 
Let us comment on the fourth one. 
Of the two possible helicity assignments to the cut propagators, one does not have
solutions for the on shell conditions. 
The other yields the coefficient of the fourth box integral in Figure~\ref{fig:5pt_1loop}:
%
%
\begin{eqnarray}
\frac{[l_2l_1]^3}{[1l_2 ][l_1 1] }
\times
\frac{\langle 2l_2\rangle^3}{\langle  2 l_3\rangle\langle l_3 l_2\rangle}
\times
\frac{[3l_4 ]^3}{[l_4 l_3] [l_3 3]}
\times
\frac{\langle l_1 l_4\rangle^3}{\langle l_4 4\rangle\langle
45\rangle\langle 5l_1\rangle }
=
- \frac{s_{12}s_{23}\;\langle
12\rangle^3}{\langle23\rangle\langle34\rangle\langle45\rangle 
\langle51\rangle}
\end{eqnarray}
The coefficients of the other integrals may be computed in a similar
fashion. They are related to the coefficient evaluated here by the obvious
relabeling the  factor $s_{12}s_{23}$. 

The quadruple cut technique described and illustrated above 
may equally well be used to construct non-planar one-loop amplitudes. 
Alternatively and perhaps less calculationally intensive, in theories with only
adjoint fields and only antisymmetric structure constant couplings  
one-loop leading and subleading color contributions are algebraically 
related \cite{Bern:1990ux} by $U(1)$ decoupling identities.

\section{Higher loops} 

Higher loop calculations in ${\cal N}=4$ sYM enjoy similar
simplifications, though to a lesser extent. An important difference
from one-loop calculations within the generalized unitarity method 
is that the natural integrals form only an over-complete basis. 
Complete bases may be identified 
on a case by case  basis\footnote{See the two-loop examples \cite{Smirnov:1999wz} 
and \cite{Gehrmann:2000zt} and the general strategy \cite{Gluza:2010ws}.}.
%
Since, in general, not all higher loop integrals can be
frozen by cutting all their propagators, a naive higher-loop 
generalization of the quadruple cuts is problematic.
The leading singularity 
method \cite{Cachazo:2008vp} bypasses the latter  difficulty by making use 
of additional propagator-like singularities in the remaining variables, which 
are specific to four dimensions.

Generalized cuts can nevertheless be used to great effect to
isolate parts of the full amplitude containing some prescribed set
of propagators. 
The previous arguments continue to hold and imply that the
complete amplitude can be reconstructed from its $d$-dimensional
generalized cuts. A detailed, general algorithm for assembling the
amplitude was described in \cite{Bern:2004cz}. In a
nutshell, starting from one (generalized) cut, one corrects it
iteratively such that all the other cuts are correctly reproduced.
%

While fundamentally all cuts are equally important, some of them
exhibit more structure than other, which makes them useful starting points for
the reconstruction of the amplitude. In some cases they also have a 
simple iterative structure and thus lead to effective rules for determining 
their contribution to the full amplitude.

\subsection{Effective rules}

Two-particle cuts are the simplest to analyze as they involve 
cutting the smallest number of propagators. For MHV amplitudes they 
exhibit special properties. As mentioned in section~\ref{susy_Ward_superspace},
to all loop orders MHV amplitudes are proportional in a natural way to 
the tree-level amplitude. 
At the level of generalized cuts this translates into the observation 
\cite{Bern:1997nh, Bern:1998ug}  that sewing two tree-level MHV
amplitudes leads in a natural way to another tree-level MHV amplitude
factor:
\begin{eqnarray}
 &&A_{n_1}^{(0),\text{MHV}}\times A_{n_2}^{(0),\text{MHV}}
\propto A^{(0),\text{MHV}}_{n_1+n_2-4}~~.
\end{eqnarray} 
%
%
The operation may be repeated, leading to what is known as "iterated 
two-particle cuts". For four-particle amplitudes,  the higher-loop
terms detected by iterated two-particle cuts are effectively given by
the rung-rule \cite{Bern:1997nh}.
It states that the $L$-loop integrals which follow
from iterated two-particle cuts can be obtained from the $(L - 1)$-loop amplitudes by
adding a rung in all possible (planar) ways while in each instance also inserting the
numerator factor 
\begin{equation}
i(l_1+l_2)^2
\end{equation}
where $l_1$ and $l_2$ are the momenta of the lines connected by the rung.
\footnote{
The rung rule can generate integrals which do not exhibit any two-particle cuts.
Such contributions must be checked by a direct evaluation of other cuts. 
Examples in this direction first appear in the planar four-loop four-gluon 
amplitude \cite{Bern:2006ew}.
%
} 
For higher-point amplitudes the rung rule is less effective and a direct 
evaluation of generalized cuts is typically necessary.

The box substitution identity \cite{Bern:2007ct} and its generalizations \cite{Bern:2010tq} 
relate further terms in higher-loop amplitudes to terms in lower loop amplitudes. 
The idea is to organize terms in an $L$-loop amplitude to expose an $L'$-loop four-point 
sub-amplitude. A contribution to the $(L+\ell)$-loop amplitude is then obtained by literally 
replacing this $L'$-loop four-point sub-amplitude with its $(L'+\ell)$-loop  counterpart.

Certain non-planar contributions to scattering amplitudes turn out to be related to 
planar ones at the same loop order  by a Jacobi-like
identities~\cite{Bern:2010tq, Bern:2008qj, Bern:2010ue}.  
   Such manipulations can be carried out pictorially.
We will not describe them in detail here, but refer the reader to the original
literature for a detailed discussion (see also \cite{BjerrumBohr:2010zs} for a string 
theory based argument for these relations).

Quite generally, effective rules do not yield all contributions to
amplitudes. Their usefulness should not, however, be underestimated:
it is easier to correct an existing ansatz rather than construct it from scratch 
starting from generalized cuts.
To determine the missing terms and confirm the ones
obtained through effective rules it is necessary to directly evaluate
certain judiciously chosen set of the generalized cuts. 

\subsection{An example: two-loop four-point amplitude in \texorpdfstring{${\cal N}=4$}{N=4} sYM theory}

Perhaps the simplest example that illustrates the higher-loop
discussion in the previous subsections is the calculation \cite{Bern:1997nh}
of the  two-loop four-point amplitude.
%
%
Direct evaluation of the $s$-channel iterated 2-particle cut (the $t$-channel 
cut may be obtained by simple relabeling) leads to:
\begin{eqnarray}
&&A_4^{(0)}{\scriptstyle (l_2, k_1^-, k_2^-, l_1)}
  A_4^{(0)}{\scriptstyle (-l_1, -l_4, -l_3, -l_2)}
  A_4^{(0)}{\scriptstyle (l_4, k_3^+, k_4^+, l_3)}\cr
&&~~~~
=i s_{12}(k_2-l_4)^2 \frac{1}{(l_2-k_1)^2(l_2+l_3)^2}A_4^{(0)}{\scriptstyle (-l_3,1^-,2^-,-l_4)}
                                                     A_4^{(0)}{\scriptstyle (l_4, k_3^+, k_4^+, l_3)}
                                                     \nonumber\\[1pt]
&&~~~~
=A^{(0)}_4{\scriptstyle (k_1^-,k_2^-,k_3^+,k_4^+)}
   \left[is_{12}(k_2-l_4)^2 \frac{1}{(l_2-k_1)^2(l_2+l_3)^2}\right]
   \left[is_{12}s_{23}\frac{1}{(l_3-k_1)^2(l_3+k_4)^2}\right]\cr
&&~~~~
=-s_{12}^2s_{23}\,A_4^{(0)}{\scriptstyle (k_1^-,k_2^-,k_3^+,k_4^+)}
\pic{15}{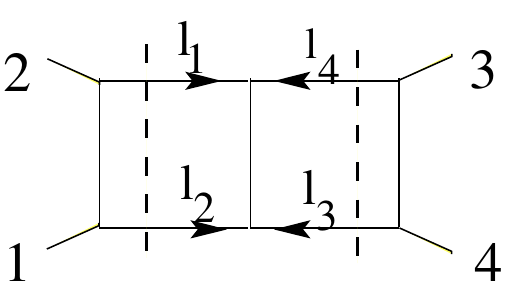}{.5}~~.
\label{4pt_D2PC}
\end{eqnarray}
Together with the loop expansion parameter (\ref{loop_exp_param}), this leads to the following ansatz:
\begin{eqnarray}
\frac{A^{(2)}_4{\scriptstyle (k_1,k_2,k_3,k_4)}}{A^{(0)}_4{\scriptstyle (k_1,k_2,k_3,k_4)}}
=-\frac{1}{4}
s_{12}s_{23}\,\left\{~s_{12} \pic{15}{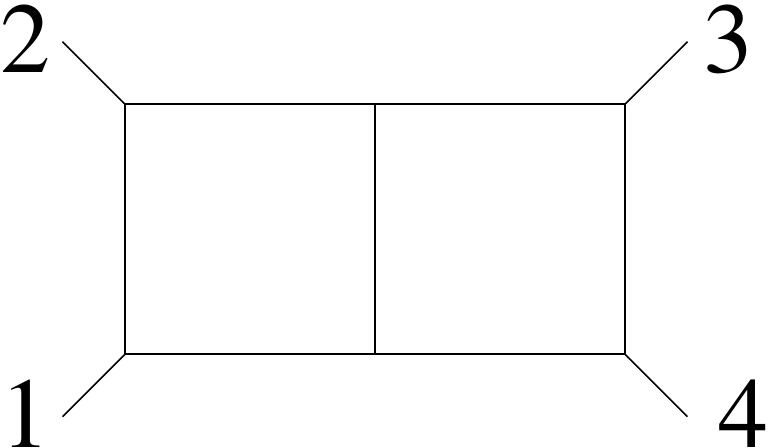}{.25}~+\;s_{23}
\pic{25}{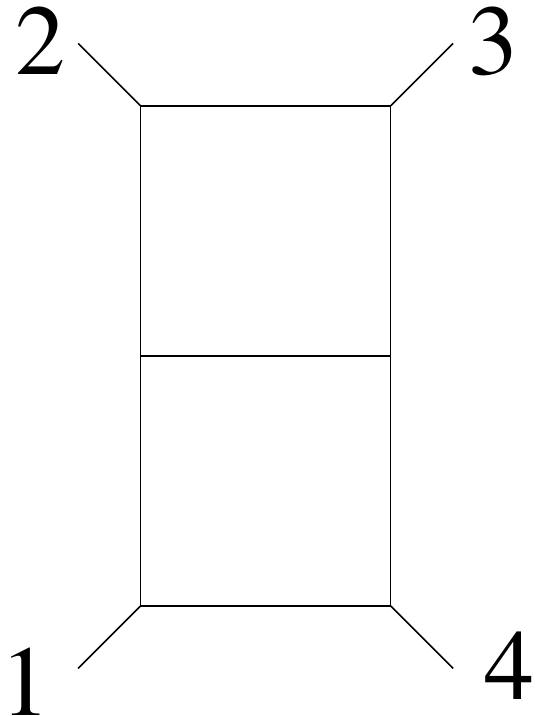}{.25}~\right\}~~.
\label{4pt2L_ans}
\end{eqnarray}
This ansatz turns out to be complete, as can be verified by evaluating
the three-particle cut \cite{Bern:1997nh}. In less supersymmetric
theories additional contributions are necessary.

The same ansatz (\ref{4pt2L_ans}), to be checked through a three-particle cut calculation, 
may be obtained either through the rung rule (by inserting a rung in the $s$- and the 
$t$-channel in a one-loop box integral) and the box insertion identity. 

In general, the evaluation of a complete (spanning) set of cuts is always 
necessary.  The power of effective rules lies in that they
provide a fast and rather effortless way of obtaining a large number
of terms (and sometimes all terms) in the amplitude. It is technically
much more convenient to test and complete an existing ansatz than to
construct it starting from the expressions of generalized unitarity
cuts.

\subsection{An interesting integral basis; dual conformal invariance}

An interesting over-complete  basis (at least for MHV amplitudes) may be
conjectured based on the observation \cite{Drummond:2006rz} that the 
integrals appearing in the two- and three-loop four-gluon planar
amplitudes exhibit a momentum space conformal symmetry known as dual
conformal symmetry.
\footnote{This symmetry,  which appears to be related to the 
higher nonlocal symmetries of the dilatation operator of  
${\cal N}=4$ sYM is reviewed in detail in \cite{chapDual}.} 
%
Curiously, this symmetry is exhibited separately by each integral 
appearing 
in the amplitude, when regularized in a specific way. In dimensional
regularization they are known as pseudo-conformal integrals. 
Dual conformal symmetry was shown to also be present in certain
higher-loops and for higher-point amplitudes;
it has been 
conjectured \cite{Bern:2006ew, Drummond:2006rz}
that, to all orders in perturbation theory, planar scattering
amplitudes exhibit this symmetry  and that each integral 
in their expressions is pseudo-conformal. Since only the infrared
regulator breaks dual conformal invariance, extraction of the known
infrared divergences (\ref{soft_collinear}) should lead yield a dual 
conformally invariant quantity.
For MHV amplitudes this conjecture applies  to 
the parity-even part of the scalar factor. 
For non-MHV amplitude it has been proposed \cite{Drummond:2008vq}  
that the ratio between the resumed amplitude and the MHV amplitude 
with the same number of external legs is invariant  under dual
conformal transformations.
This conjecture was successfully tested for the (appropriately defined) even 
part of the six-point NMHV amplitude at two loops~\cite{Kosower:2010yk}.

The even part of planar MHV amplitudes
is expected to be a sum of pseudo-confomal
integrals
with constant coefficients: 
\begin{eqnarray}
{\cal M}_n^{(L)}=\sum_i \; c_i \, I_i~~;
\label{ansatz_DC}
\end{eqnarray}
the coefficients $c_i$ may be determined by comparing cuts of this
ansatz to direct evaluation of generalized cuts of the amplitude. In
certain cases maximal cuts are sufficient.  This strategy was used to
determine the five-loop four-point amplitude \cite{Bern:2007ct}
as well as the two-loop MHV amplitudes with any number of external 
legs \cite{Vergu:2009tu}.

\section{\label{other} Comments on other methods and outlook}

Other methods have been put forward for the construction of scattering
amplitudes in ${\cal N}=4$ sYM theory and, more generally, in
maximally-supersymmetric theories. A notable one, which captures the
spirit of the complete localization of one-loop integrals under
quadruple cuts, is the so-called leading singularity conjecture
\cite{ArkaniHamed:2008gz}. As previously discussed, evaluating the
maximal cuts of an amplitude does not lead to a complete localization
of integrals. In certain cases the result however exhibits further
propagator-like singularities which may also be cut. The result is
known as the "leading singularity". The conjecture states that
scattering amplitudes in maximally supersymmetric theories are
completely determined by their leading singularities. 
Two-loop results based on this conjecture agree with the results of
the unitarity method calculation.  It was also used to construct
\cite{Cachazo:2008hp} the odd part of the six-point MHV amplitude at
two-loops as well as \cite{Spradlin:2008uu} the three-loop five-gluon
amplitude. Together with the assumption that the superconformal and 
dual superconformal symmetries are realized to all orders in perturbation 
theory, it led to a proposal \cite{ArkaniHamed:2010kv} for the all-loop 
all-point planar scattering amplitudes of the ${\cal N}=4$ sYM theory;
a specific regularization prescription is required.
The six-point MHV amplitude is 
correctly reproduced by this proposal \cite{Drummond:2010mb};  this 
calculation also emphasizes that, in this proposal, the natural integrals  
are technically simpler than standard Feynman integrals. It 
has been suggested that this is related to them having {\it only} unit leading 
singularities.

All-order expressions of scattering amplitudes are in general hard to
construct. Based on explicit two-loop \cite{Anastasiou:2003kj} and
three-loop calculations \cite{Bern:2005iz} as well as on the collinear
properties of amplitudes it was conjectured that the scalar factor of
$n$-point MHV amplitudes has, to all loop orders, a simple iterative
structure in terms of the corresponding one-loop amplitude
\cite{Bern:2005iz} to all orders in $d=4-2\epsilon$ dimensions:
\begin{eqnarray}
{\cal M}_n=
\exp\left[\sum_{l=1}^\infty \, a^l f^{(l)}(\epsilon){\cal M}^{(1)}_n(l\epsilon)+C^{(l)}
+{\cal O}(\epsilon)\right]
\label{exp_all_n}
\end{eqnarray}
where $f^{(l)}(\epsilon)=f_0^{(l)}+\epsilon f_1^{(l)}+\epsilon^2f_2^{(l)}$ with $f_0^{(l)}$ and 
$f_1^{(l)}$ determined in terms of the similar coefficients appearing in the Sudakov form 
factor (\ref{soft_collinear}),~(\ref{coefs}).

For $n=4,5$ this expression appears to hold \cite{Drummond:2007cf} 
if dual conformal invariance is present to all orders in perturbation theory. 
At higher-points dual conformal invariance is no
longer sufficient to fix the expression of the amplitude.  Direct
calculations \cite{Bern:2008ap} of the six-point amplitudes show a
departure from this expression, initially anticipated from a strong
coupling analysis \cite{Alday:2007he} based on the proposed relation 
between  planar MHV scattering amplitudes and certain null polygonal Wilson 
loops \cite{Alday:2007hr} in this regime (see also \cite{chapTDual}).  
The so-called ``remainder function" quantifies this difference; its analytic form
was found in \cite{DelDuca:2009au} and simplified in \cite{Goncharov:2010jf}.

The proposed relation between planar MHV scattering amplitudes and 
Wilson loops \cite{Alday:2007hr} led to the conjecture 
\cite{Drummond:2007cf, Drummond:2007aua, Brandhuber:2007yx} 
that a similar relation may holds order by order in weak coupling perturbation theory. 
This topic is reviewed in detail in \cite{chapDual}.
The comparison of the six-point MHV amplitude at two loops with the relevant Wilson 
loop was discussed in \cite{Bern:2008ap, Drummond:2008aq}. Expectation values of Wilson 
loops relevant for higher-point amplitudes have been computed in \cite{Anastasiou:2009kna};
comparison with the corresponding scattering amplitude 
calculations \cite{Vergu:2009tu, Vergu:2009zm} awaits further developments in the 
calculation of higher-loop higher-point Feynman integrals.

Throughout our discussion we assumed that IR divergences are
regularized in dimensional regularization. Ultraviolet
divergences not being an issue in ${\cal N}=4$ sYM, infrared
divergences may also be regularized by letting fields acquire masses
through spontaneous breaking of the gauge symmetry
\cite{Alday:2009zm} (Higgs regularization). 
Much like the original (all-massive) regularization of
\cite{Drummond:2006rz}, this regularization has the advantage of
preserving dual conformal invariance up to transformation of the mass
parameter(s). This regularization was used to great effect to test the
exponentiation (\ref{exp_all_n}) of the four-point amplitude at two- and three-loops
\cite{Alday:2009zm, Henn:2010bk}. 
Diagrammatic rules, based on the color flow, may be
devised to avoid repeating the unitarity-based construction in the
presence of mass parameters. 
The Higgs-regularized amplitude may also be obtained from the dimensionally
regularized one by simply treating as mass parameters 
the $(-2\epsilon)$-dimensional components of loop momenta. 
%
%
A calculation is necessary to ascertain whether the Higgs-regularized amplitude 
contains terms proportional to the regulator which yield non-vanishing contributions 
upon integration.

Being somewhat outside the main theme of the collection, 
we glossed over the very important techniques developed
specifically for the calculation of nonplanar scattering amplitudes, in particular the 
Bern-Carrasco-Johansson relations \cite{Bern:2008qj, Bern:2010ue} and the connection
between ${\cal N}=4$ sYM theory and ${\cal N}=8$ supergravity.

The full consequences and implications of the developments outlined in
this review (as well as of those that were not) are yet to emerge and
many questions, which will undoubtedly contribute in this direction,
remain to be addressed. Despite substantial progress in the
calculation of multi-loop and multi-leg amplitudes there is room for
improvement. It is clear that further 
structure is present in ${\cal N} = 4$ sYM theory and that it may be sufficiently
powerful to completely determine, at least in some sectors, the
kinematic dependence of the scattering matrix of the theory.

\section*{Acknowledgements}
I would like to thank  Z.~Bern, J.J.M.~Carrasco, L.~Dixon, 
H.~Johansson, D.~Kosower, M.~Spradlin, A.~Tseytlin, C.~Vergu 
and A~.Volovich for many useful discussions and collaboration on the 
topics reviewed here.
This work was supported by the US DoE under contract
DE-FG02-201390ER40577 (OJI), 
the US NSF under grants PHY-0608114 and 
PHY-0855356 and the A.\ P.\ Sloan Foundation.


\phantomsection
\addcontentsline{toc}{section}{\refname}
\bibliography{intads,AmplitudesBibliography,chapters}
\bibliographystyle{nb}

\end{document}